\documentclass{aa}  

\usepackage[acronym]{glossaries}
\usepackage{CJKutf8}
\usepackage{graphicx}
\usepackage{CJK}
\usepackage{subcaption}
\usepackage{color}
\usepackage{ulem}
\usepackage{booktabs}
\usepackage{caption}
\usepackage{float}
\usepackage{natbib,twoopt}
\usepackage[breaklinks=true]{hyperref} 
\hypersetup{
    colorlinks=true,
    linkcolor=blue,
    filecolor=blue,      
    urlcolor=blue,
    citecolor=blue,
}
\usepackage{caption}
\usepackage{txfonts}

\newacronym{GW}{GW}{gravitational wave}
\newacronym{IMBH}{IMBH}{intermediate-mass black hole}
\newacronym{IMBBH}{IMBBH}{intermediate-mass binary black hole}
\newacronym{BBH}{BBH}{binary black hole}
\newacronym{SMBH}{SMBH}{supermassive black hole}
\newacronym{SBH}{SBH}{stellar mass black hole}
\newacronym{SBBH}{SBBH}{stellar mass binary black hole}
\newacronym{BH}{BH}{black hole}
\newacronym{AGN}{AGN}{active galactic nuclei}
\newacronym{Pop III}{Pop III}{population III}
\newacronym{EM}{EM}{electromagnetic}
\newacronym{LVK}{LVK}{LIGO/Virgo/KAGRA}
\newacronym{ET}{ET}{Einstein Telescope}
\newacronym{CE}{CE}{Cosmic Explorer}
\newacronym{PTA}{PTA}{pulsar timing array}
\newacronym{S/N}{S/N}{signal-to-noise ratio}
\newacronym{VMS}{VMS}{very massive star}

\makeglossaries

\begin{document} 
\begin{CJK*}{UTF8}{gbsn}

\title{Detection for intermediate-mass binary black holes in population III star clusters with TianQin}

\subtitle{}

\author{Hanzhang Wang
          \inst{1}
          \and
          Shuai Liu
\inst{2}\thanks{Corresponding author: liushuai@zqu.edu.cn}
          \and
          Han Wang
          \inst{1}
          \and
          Hong-Yu Chen
          \inst{1}
          \and
          Long Wang
          \inst{3}
          \and
          Yiming Hu
          \inst{1}\thanks{Corresponding author: huyiming@sysu.edu.cn}
          }

   \institute{MOE Key Laboratory of TianQin Mission, TianQin Research Center for Gravitational Physics \& School of Physics and Astronomy, Frontiers Science Center for TianQin, Gravitational Wave Research Center of CNSA, Sun Yat-sen University (Zhuhai Campus), Zhuhai, 519082, China
         \and
             School of Electronic and Electrical Engineering, Zhaoqing University, Zhaoqing 526061, China
        \and
             School of Physics and Astronomy, Sun Yat-sen University (Zhuhai Campus), Zhuhai 519082, China
             }

   \date{Received XXXX; accepted YYYY}

  \abstract
   {Population III star clusters are predicted to form in unenriched dark matter halos.
   Direct N-body simulation of Pop III clusters implies the possible formation and merger of intermediate-mass binary black holes (IMBBHs). The gravitational wave signals could be detected by space-borne gravitational wave detectors such as TianQin.}
   {This study evaluates the potential of  TianQin in detecting IMBBHs from Pop III star clusters, focusing on key factors such as the detection horizon, detection number, and Fisher uncertainty.}
   {A Monte Carlo simulation is employed to derive IMBBH sources, utilizing the IMRPhenomD waveform model to simulate catalogs of observable IMBBH mergers. The mass and redshift distributions are derived from direct N-body simulations of IMBBHs in Population III star clusters. Detection numbers are determined by calculating the signal-to-noise ratios (S/N) of the simulated signals and applying thresholds for detection. Fisher uncertainty is obtained through Fisher information matrix analysis.}
   {The findings suggest that TianQin could achieve detection numbers within 5 years ranging from 1 in the most pessimistic scenario to 253 in the most optimistic scenario. Furthermore, TianQin can precisely constrain the IMBBH mass with a relative uncertainty of $10^{-6}$, coalescence time $t_c$ within 1 second, and sky location $\bar{\Omega}_S$ within 1 $\rm{deg}^2$. However, the luminosity distance $D_L$ and inclination angle $\iota$ exhibit substantial degeneracies, limiting their precise estimation.}
   {}


   \maketitle
%
\section{Introduction}
\Glspl{IMBH}, which possess masses ranging from $10^{2}\text{M}_\odot$ to $10^5 \, \text{M}_\odot$ \citep{miller04}, are considered to be the link between \glspl{SMBH} and \glspl{SBH}. 
The search for them is a prominent topic, and the detailed study of the \gls{IMBH} demographics can help answer key questions such as the seeding mechanisms of \glspl{SMBH} \citep{greene20}, and the co-evolution of galaxies and their central black holes \citep{volonteri10,kormendy13,king23}.

The efforts to detect \glspl{IMBH} through EM observations have not produced the definitive evidence of their existence, \citep[e.g.,][]{rees78,colbert06,miller04,koliopanos18,greene12,mezcua17}, although several candidates have been proposed within globular clusters, such as 47 Tucanae \citep{kiziltan17} and Omega Centauri \citep{baumgardt17}, as well as in certain dwarf galaxies, exemplified by 3XMM J215022.4-055108 \citep{lin18, lin20, wen21}. Furthermore, some ultraluminous X-ray sources (ULXs), including M82 X-1 and HLX-1, are also regarded to host potential \glspl{IMBH} \citep{ebisuzaki01, kaaret01, matsumoto01, miller02, hopman04, portegies04, king05, patruno06, pasham13}. However, EM observations cannot confirm whether a signal originates from an IMBH or multiple black holes \citep{pooley07}. On the other hand, the \gls{GW} observation from ground-based detectors such as LIGO and Virgo result in several convincing detections of \glspl{IMBH}, including the merger event GW190426\_190642, GW190521, and GW200220\_061928 \citep{abbott24,abbott23, abbott20}. The deduced masses of the remnant black holes all excee 100 $M_\odot$. 

On the theoretical side, there are also challenges in understanding the formation channels and sites of \glspl{IMBH}. Several formation mechanisms have been proposed in addition to the binary black hole mergers, which can be categorized into three principal approaches \citep{greene20}: the collapse of \gls{Pop III} stars \citep{madau01,fryer01,bromm02, bromm04,lodato06, begelman06, ryu16}, the direct collapse of gas \citep{loeb94,bromm03, latif13,shi24}, and gravitational runaway mergers involving stars and \glspl{BH} within dense star clusters \citep{gonzalez21, rizzuto21,kritos23, kritos24b, kritos24a, kritos24c, sharma25, fujii24}.  
In the early universe, the \gls{Pop III} stars are expected to have been very massive because molecular hydrogen cooling was inefficient \citep{bromm04, karlsson13}. These stars likely produce black hole remnants of about $100M_\odot$, except for those in the $140-260M_\odot$ range, which instead explode as pair-instability supernovae and left no remnant \citep{fryer01, heger03}. \glspl{IMBH} ranging from $10^3-10^4\text{M}_\odot$ can be formed through gravitational runaway mergers\citep{miller02, portegies04}. Additionally, direct collapse can result in \glspl{IMBH} of $10^4-10^6\text{M}_\odot$, avoiding gas fragmentation due to the suppression of cooling \citep{visbal14, habouzit16} and collapses directly into a black hole. 
For the formation time, \gls{Pop III} star collapse and direct collapse predominantly occur at high redshifts, while gravitational runaway mergers can occur at all redshifts, depending on the age and dynamics of the star clusters. 

To date, there have been no direct \gls{GW} detections of  \glspl{IMBH} mergers. 
This is not surprising as the current ground-based \gls{GW} facilities are not optimized for detecting \glspl{BBH} including \glspl{IMBH} (IMBH-BH) \citep{han17}. Compared with \gls{LVK}, the upcoming space-borne \gls{GW} observatories such as TianQin \citep{luo16} and LISA \citep{danzmann96}, and the next generation ground-based \gls{GW} detectors such as \gls{ET} \citep{punturo10} and \gls{CE} \citep{reitze19}, are more sensitive at the lower frequency ranges, thus can detect binaries with \glspl{IMBH} more efficiently. 
Theoretical investigations have further underscored the potential of future \gls{GW} detectors in exploring \glspl{IMBH}. For instance, \cite{arca21} examined intermediate-mass ratio inspirals (IMRIs) involving \glspl{IMBH} and compact stellar objects, offering valuable insights into their detection capabilities. \cite{rasskazov20} conducted simulations regarding the formation of \glspl{IMBH} within globular clusters, predicting that LISA could detect at least one merger event during a four-year mission. 
Moreover, \cite{fragione18a, fragione18b, fragione20} developed cluster models that incorporated \glspl{IMBH}, demonstrating that observatories such as \gls{ET}, \gls{CE} and LISA could potentially identify IMBH-BH mergers across a broad mass spectrum.
Furthermore, other observational techniques have also shown promise in detecting \glspl{IMBH}. For instance, \cite{strokov22, strokov23} investigated LISA’s ability to detect and constrain IMBHs indirectly via the Doppler effect. \cite{torres23} studied the capability of atom interferometers such as DECIGO \citep{kawamura11} to detect \glspl{IMBH}. Gravitational lensing has also been proposed as a viable method for detecting \glspl{IMBH} \citep{kains16}, while \glspl{PTA} has been identified as another promising tool for detecting \glspl{IMBH} through prolonged \gls{GW} observations \citep{kocsis12,sesana14, dror19, barausse23, steinle23}. 

\begin{figure}[hbt!]
    \centering
    \includegraphics[width=0.6\linewidth]{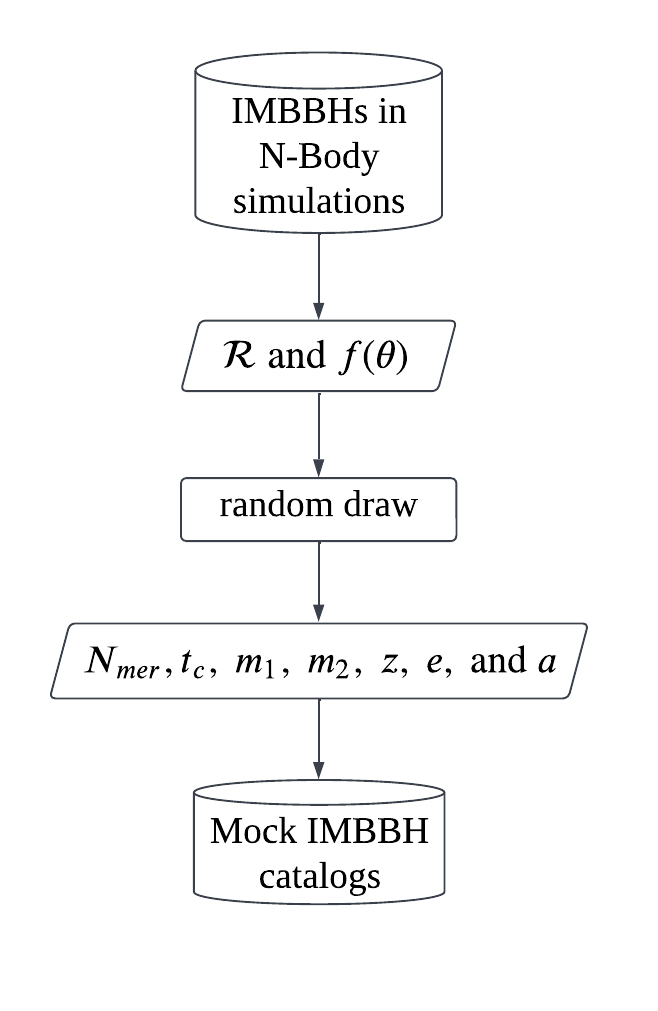}
    \caption{Process of generating mock IMBBH catalogs. $\mathcal{R}$ denotes the merger rate, while $\theta$ represents IMBBH parameters, including component masses $m_1$ and $m_2$, redshift $z$, eccentricity $e$, and semi-major axis $a$. The distribution function is denoted as $f(\theta)$. }
    \label{fig:IMBBH-flow}
\end{figure}

TianQin is an mHz space-borne \gls{GW} observatory proposed by China, which would be launched in the 2030s \citep{luo16, hu18, mei21}. TianQin has an equilateral triangular configuration composed of three satellites orbiting the Earth at an approximate distance of $10^5 \, \text{km}$. Compared with LISA, TianQin has shorter arms, which makes it more sensitive to the lighter end of the IMBH mass spectrum \citep{hu24,li25}. 
\cite{wang22} and \cite{liu24} explored the \gls{IMBH} formation processes and their population properties in \gls{Pop III} clusters, but the TianQin detection potential for these systems have not been studied in detail. In this work, we aim to quantitatively explore TianQin's detection ability for \glspl{IMBBH} formed within \gls{Pop III} clusters. As Fig.\ref{fig:IMBBH-flow} shows, we first derived the evolution history of \glspl{IMBBH} formed in the \gls{Pop III} clusters based on direct N-body simulations. Combining the merger history and parameter distributions with the spatial distribution of \gls{Pop III} clusters, we then obtained a mock catalog of \gls{IMBBH} mergers. By calculating the \gls{S/N} with TianQin and applying the Fisher information matrix (FIM), we then deduced the expected detection number and Fisher uncertainty.

The rest of this paper is organized as follows. We investigate the population properties of \glspl{IMBBH} and introduce the methods used in Sec. \ref{sec:method}. In Sec. \ref{sec:result}, we estimate the detection capacity of TianQin for IMBBHs. Finally, we draw the conclusion in Sec. \ref{sec:conclusion}. Throughout the paper, we adopt the geometrical unit, i.e., $G=c=1$, and the standard $\Lambda$CDM cosmological model \citep{planck16}.
\section{Method}\label{sec:method}
In this section, we introduce the N-body simulation approach and the initial conditions we use to trace the evolution of {Pop III} clusters embedded in mini dark matter halos. Furthermore, we analyze the population features of \glspl{IMBBH} from N-body simulation and introduce the methods utilized to estimate the detection capacity of TianQin for \glspl{IMBBH}, including \gls{GW} response signal, \gls{S/N}, and FIM.

\subsection{N-body simulation}

In previous studies, \cite{sakurai17} investigated IMBH formation without performing long-term simulations, while \cite{reinoso18} examined low-mass star-cluster models with minimum Z=0.0001. In addition, \cite{wang22} studied the long-term evolution of \gls{Pop III} clusters and their IMBHs but did not include primordial binaries. 

In this work, we simulated the evolution of \gls{Pop III} star clusters using the N-body code PETAR \citep{wang20} coupled with the single and binary population-synthesis code BSEEMP \citep{tanikawa20}.
The initial conditions for the clusters in this study are derived from the long-term model \texttt{NFWden\_long\_w9\_noms\_imf1}, which excludes primordial binaries \citep{wang20}. This model was chosen because it predicts the formation of \glspl{VMS} that evolve into \glspl{IMBH} with masses up to $10^3 M_\odot$. Additionally, new models were introduced to include primordial binaries, where the initial periods, mass ratios, and eccentricities follow the distributions from \cite{sana12}.

Key parameters for the simulation of \gls{Pop III} clusters are as follows. The initial mass of the star clusters $ M_{\text{clu}} =10^5 M_\odot$ \citep{sakurai17}, with a half-mass radius $ r_h = 1  \text{pc}$, which is a typical value for observed star clusters. The central density profile of the clusters is described using the Michie-King model \citep{michie62, king66}, where the cluster compactness is determined by the ratio of the core radius $ r_c $ to the tidal radius $ r_t $, denoted as $ W = r_c/r_t$. For our study, we set the initial value of $ W_0 = 9 $.
Furthermore, the initial mass function (IMF) for the stars in these Pop III clusters follows a top-heavy distribution, based on the results from hydrodynamic simulations \citep{stacy16, chon21, latif22}. Specifically, the IMF is described by a single power-law profile
$$
p(m) \propto m^{-1}, \quad 1M_\odot < m < 150M_\odot,
$$
which favors the formation of \glspl{VMS} and \glspl{BH}, in contrast to the canonical IMF characterized with a power index of $ -2.35 $ \citep{kroupa01, chabrier03}.
Moreover, for the dark matter halo surrounding the Pop III clusters, we assume that its potential follows the Navarro-Frenk-White (NFW) profile \citep{navarro96}, given by

$$
\Phi = -\frac{G M_{\text{vir}}}{r \left[ \log(1 + C) - C/(1 + C)\right] }\log\left( 1 + r/r_{\text{vir}}\right),
$$
where $ M_{\text{vir}} = 4 \times 10^7 M_\odot $ is the virial mass, $ r_{\text{vir}} = 280 \, \text{pc} $ is the virial radius, and the concentration $ C(z) $ evolves as $ C(z) = C(0) / (1 + z) $, where $ C(0) = 15.3 $ is the concentration value for the Milky Way halo \citep{bovy15}.

\subsection{Population of IMBBH}\label{sec:population}

\begin{figure}[hbt!]
\centering
\includegraphics[width=0.49\textwidth]{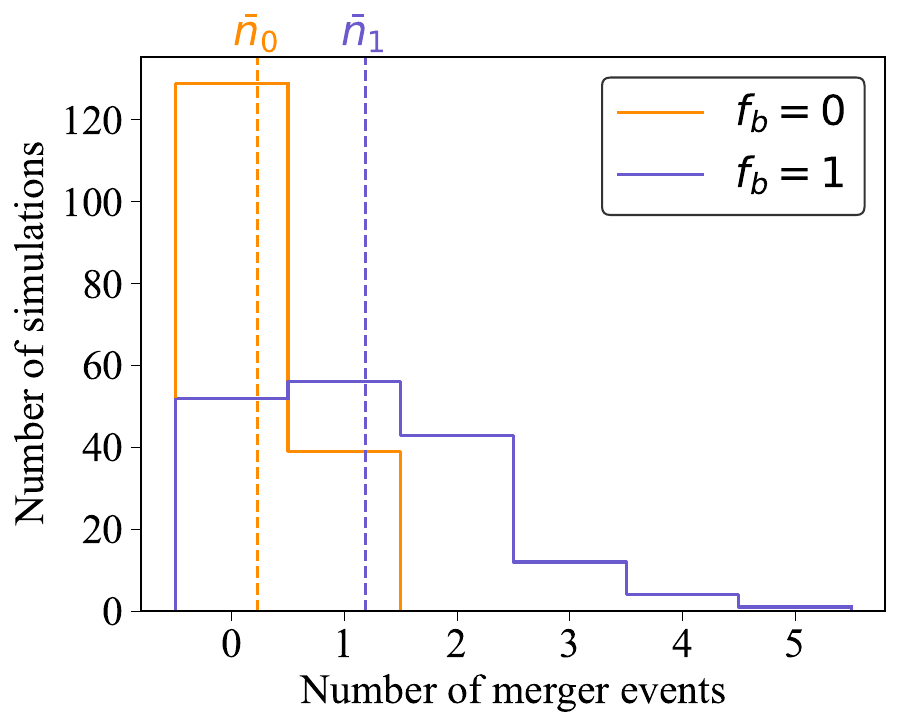}
\caption{Distribution of the merger numbers of \glspl{IMBBH} from 168 N-body simulations. Solid lines represent $f_{\rm b}=0$ (orange) and $f_{\rm b}=1$  (blue), respectively. $\bar{n}_0 \approx 0.2$ and $\bar{n}_1 \approx 1.2$ are average numbers of IMBBH mergers for the two scenarios. }
\label{fig:ori_IMBBH_num}
\end{figure}

We adopted the N-body simulation results of Pop III cluster model `NFWden\_long\_w9\_noms\_imf1' conducted by \cite{liu24}, where 168 Pop III clusters with primordial binary fraction $f_{\rm b}=0$ and 1 were evolved for up to 12Gyr, respectively. We selected the \gls{IMBBH} mergers formed in these simulations and explored their population properties as below. The distribution of \gls{IMBBH} merger numbers is shown in Fig. \ref{fig:ori_IMBBH_num}.

When $f_{\rm b}=0$, in most simulations no IMBBH would merge, while for the remaining cases, one pair of IMBBH could merge. As $f_{\rm b}$ increases to 1, most simulations have at least one IMBBH merger, with a maximum of five IMBBH mergers. This indicates that clusters with a higher primordial binary fraction ($f_b =1$) can produce more \gls{IMBBH} events than their lower primordial binary fraction counterparts ($f_b =0$).

\begin{figure*}[hbt!]
\centering
\includegraphics[width=\linewidth]{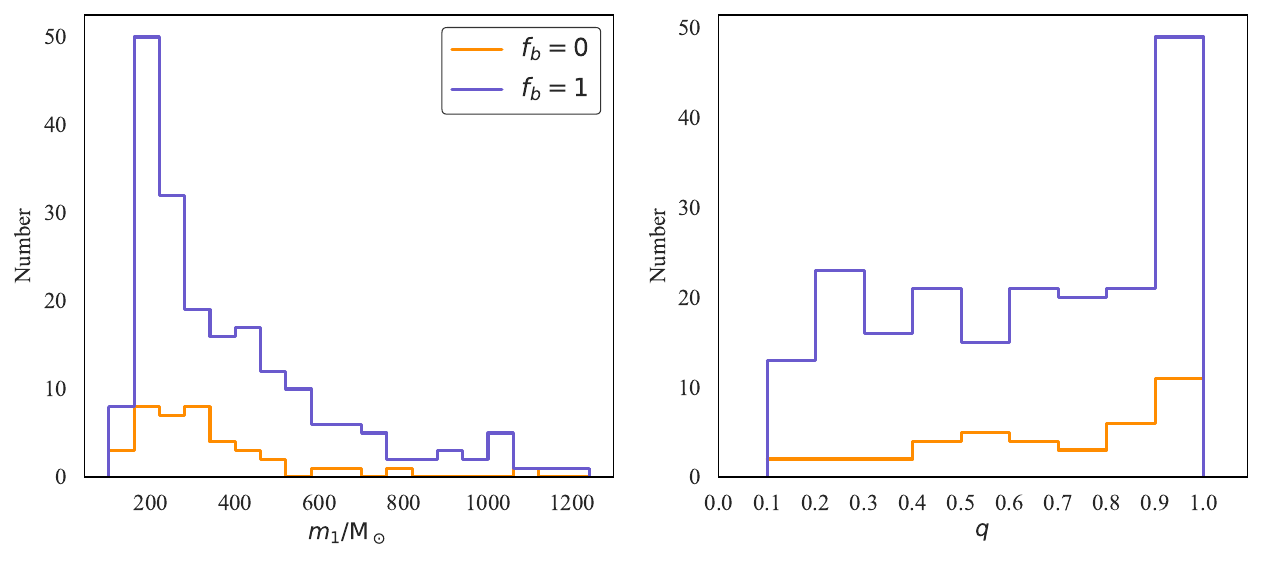}
\caption{Distributions of primary mass $m_{1}$ (left panel) and mass ratio $q=m_{2}/m_{1}$ (right panel) of \glspl{IMBBH} from the \gls{Pop III} cluster models with $f_{\rm b}=0$ (orange) and $f_{\rm b}=1$ (blue). }
\label{fig:mass distribution}
\end{figure*}

The distributions of \glspl{IMBBH} mass parameters are presented in Fig. \ref{fig:mass distribution}. In both cases, the primary masses $m_{1}$ peaks around $200M_{\odot}$, with a decreasing tail extends to $\sim1000M_{\odot}$, and the $f_{\rm b}=1$ case exhibits a narrower peak. 
The distribution of mass ratio $q=m_{2}/m_{1}$ peaks at about 1, indicating that the component masses of most \glspl{IMBBH} are comparable.

\begin{figure}[hbt!]
\centering
\includegraphics[width=0.49\textwidth]{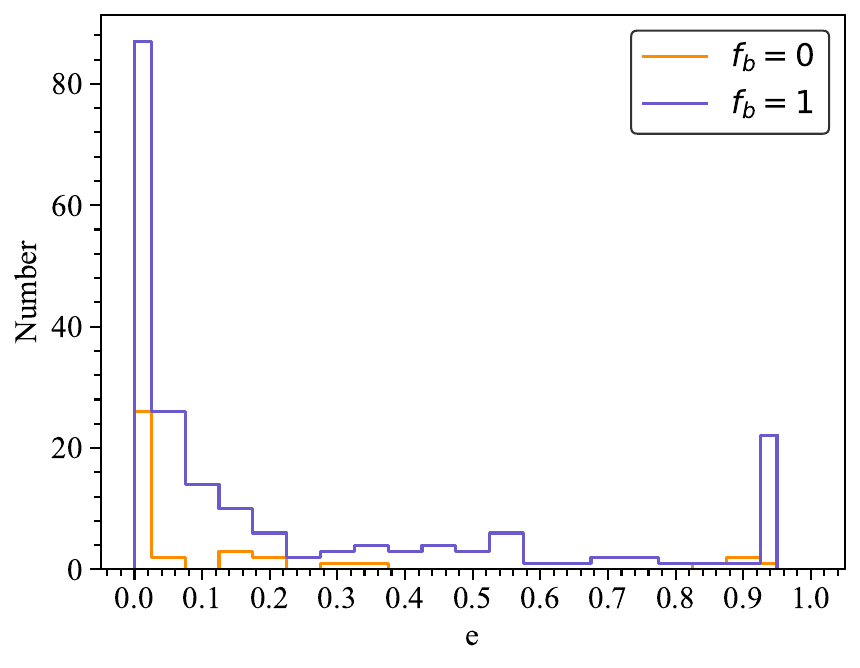}
\caption{Orbital eccentricity distribution of \glspl{IMBBH} from the \gls{Pop III} cluster models with $f_{\rm b}=0$ (orange) and $f_{\rm b}=1$ (blue) at their formation time.}
\label{fig:eccentricity}
\end{figure}

We present the distribution of eccentricities in Fig. \ref{fig:eccentricity}. In both cases, the simulations predict relatively small natal eccentricities, with approximately 70-80\% of the eccentricities in the range $e \in [0, 0.25]$. Only a small fraction of \glspl{IMBBH} have a very large eccentricity of $0.8<e<1$. Orbital decay can be accelerated through the binary–single encounter where a binary encounters a third body, the third body can replace a component in the binary. Such binary–single encounters are often associated with a shrunk orbital separation and a large orbital eccentricity for the new binary. Due to the higher portion of primordial binaries, the case with $f_{\rm b}=1$ is expected to experience more frequent binary–single encounters, resulting in a higher proportion of larger orbital eccentricity sources.

\begin{figure}[hbt!]
\centering
\includegraphics[width=0.49\textwidth]{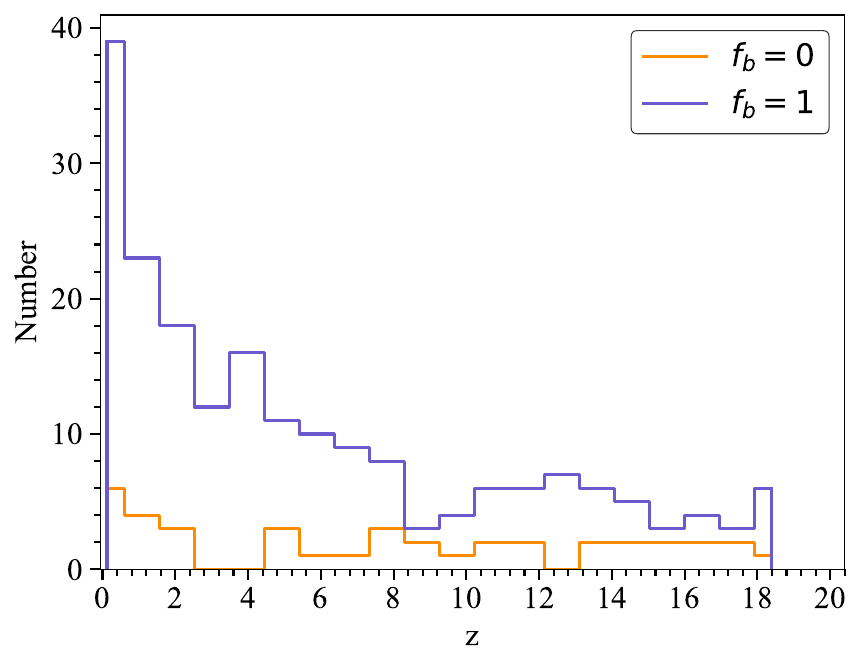}
\caption{Redshift distribution of \glspl{IMBBH}  from the \gls{Pop III} cluster models with $f_{\rm b}=0$ (orange) and $f_{\rm b}=1$ (blue) corresponding to their merger time.}
\label{fig:redshift}
\end{figure}

The redshift distributions of \glspl{IMBBH} mergers in the simulations are shown in Fig. \ref{fig:redshift}. In both cases of $f_{\rm b}=0$ and 1, approximately 40-60\% of the mergers occur at high redshifts with $ z > 6 $, and the minimum redshift $ z_{\text{min}} > 0.1 $. 

The average number of \glspl{IMBBH} per cluster is 
$N_{\text{IMBBH}}=\sum_{i=1}^{N_{sim}}{{N_\text{IMBBH}}^i}/N_{sim}$, where $N_{sim}=168$ is the total number of simulation and ${N_\text{IMBBH}}^i$ denotes the \glspl{IMBBH} merger number of each simulation. The result is 0.2 and 1.2 per simulation for $f_{\rm b}=0$ and $f_{\rm b}=1$ respectively, with the distribution of \glspl{IMBBH} merger numbers shown in Fig.\ref{fig:ori_IMBBH_num}. 

In order to extrapolate the merger number within a cluster to the cosmic merger rate, one needs to also derive the number density of Pop III clusters.
The typical mass of a Pop III cluster is $M_{sc} = 10^5 \, \text{M}_{\odot}$\citep{wang22}. 
Meanwhile, it is derived that the average stellar mass density that originates from Pop III clusters  $\phi$ ranges from $3.2 \times 10^4 \, \text{M}_{\odot} \, \text{Mpc}^{-3}$ \citep{skinner20} to $2 \times 10^5 \, \text{M}_{\odot} \, \text{Mpc}^{-3}$ \citep{inayoshi21}. 
We can then calculate the number density of Pop III clusters, $n_{\text{Pop III}}$, as
$$
n_{\text{Pop III}} = \frac{\phi}{M_{sc}}.
$$
Substituting the values of $\phi$ and $M_{sc}$, we obtained the following range for the number density of Pop III clusters, from $0.32 \, \text{Mpc}^{-3}$ to $2 \, \text{Mpc}^{-3}$. Multiplying this number density by the average number of IMBBH mergers per cluster, $N_{\text{IMBBH}}$, gives the number density of IMBBHs mergers, $n_{\text{IMBBH}}$
$$
n_{\text{IMBBH}} = n_{\text{Pop III}} \times N_{\text{IMBBH}}.
$$
The merger rate $\mathcal{R}$ of \glspl{IMBBH} averaged over redshift can be estimated \citep[see][Eq.~7]{liu24} as follows, given by the merger number density divided by the simulation time $t$
$$\mathcal{R} = \frac{n_{\text{IMBBH}}}{t}.$$
With a simulation time span of $t = 12 \, \text{Gyr}$, the merger rate is shown in Tab. \ref{tab:merger_rate}.

\begin{table}[ht]
\centering
\begin{tabular}{c l l }
\toprule
$\mathcal{R}(\text{yr}^{-1} \, \text{Gpc}^{-3})$ & $f_{\rm b} = 0$ & $f_{\rm b} = 1$ \\
\toprule
$\mathcal{R}_{lower}$ & 0.006 & 0.03 \\
\midrule
$\mathcal{R}_{upper}$ & 0.03 & 0.18 \\
\bottomrule
\end{tabular}
\caption{Lower and upper limits of average IMBBH merger rate for $f_{\rm b} = 0$ and $f_{\rm b} = 1$, denoted as $\mathcal{R}_{lower}$ and $\mathcal{R}_{upper}$.}
\label{tab:merger_rate}
\end{table}

Based on the averaged merger rate and the distributions of masses and redshift yielded by kernel density estimation, assuming a uniform distribution in source location as well as orbital orientation, described by parameters $\bar{\theta}_{S}$, $\bar{\phi}_{S}$, $\bar{\theta}_{L}$ and $\bar{\phi}_{L}$, we obtained a simulated \gls{IMBBH} samples through the Monte Carlo method. 

\subsection{Gravitational wave signal}
According to the different calculation methods, the \gls{GW} waveform of a BBH can be roughly categorized into three stages: the inspiral, the merger, and the ringdown (IMR). The \glspl{GW} from circular BBHs only contain the $n=2$ harmonic, and more harmonics are included in the case where BBHs are eccentric. For BBHs with small or moderate $e$, the $n=2$ harmonic is always dominant, whereas it no longer dominates in the case where BBHs are highly eccentric \citep{liu22,chen17}. Correctly describing the effect of orbital eccentricities can be challenging when the eccentricities are very large, especially when trying to include the whole IMR stages in the waveform.
However, the majority of simulated \glspl{IMBBH} have $e<0.3$ at formation, as illustrated in Fig. \ref{fig:eccentricity}. With circularization caused by gravitational radiation, the eccentricities shrink significantly.
Therefore, for the sake of convenience, we use the widely adopted IMRPhenomD waveform \citep{husa16,khan16} to perform the following calculation. It is a frequency domain waveform that can describe the IMR evolution of a non-precessing non-eccentric \gls{BBH}. 

\begin{figure*}[hbt!]
\centering
\includegraphics[width=\textwidth]{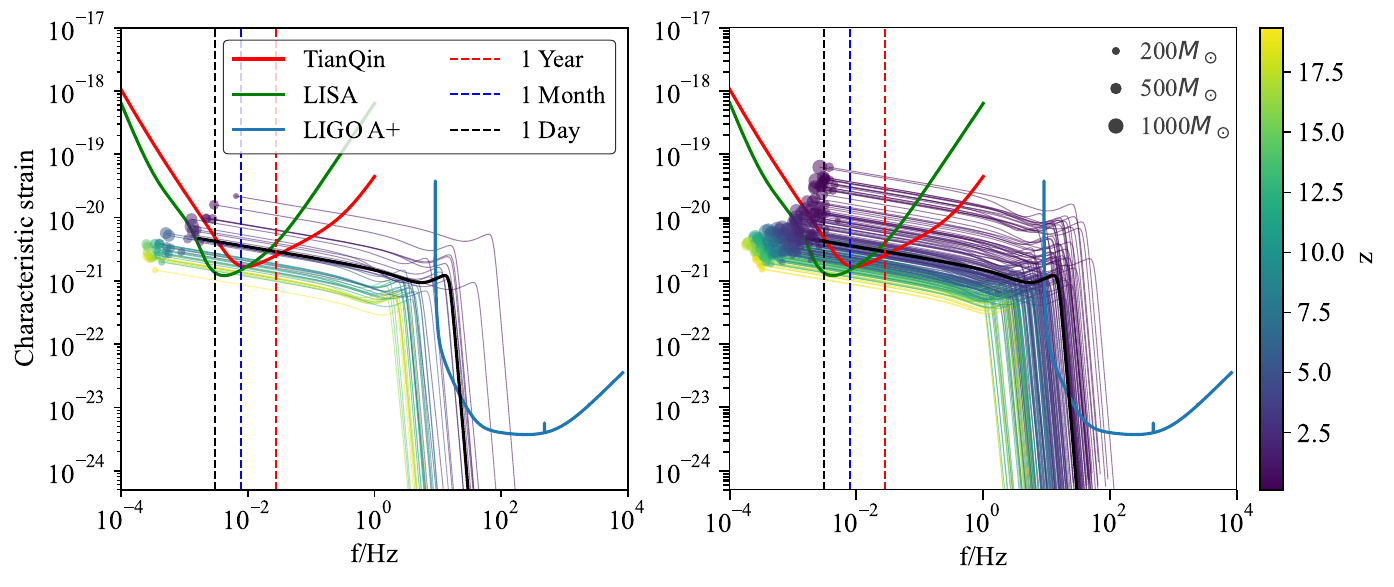}
\caption{Sky-polarization-averaged GW signals of IMBBHs modeled with IMRPhenomD (thin lines), overlaid with TianQin (red), LISA (green), and LIGO A+ (blue) sensitivity curves. The merger time is set to 5 years. Dots size scale with total IMBBH mass. Left and right panels correspond to $f_{\rm b} = 0$ and $f_{\rm b} = 1$, respectively. The black solid line represents a reference IMBBH ($m_1 = m_2 = 200,M_\odot$, $z = 2$, $t_c = 5$ yr), with red, blue, and black dashed lines marking 1 year, 1 month, and 1 day before merger.}
\label{fig:gw signal}
\end{figure*}

When \glspl{GW} from BBHs reach detectors, their waveforms are modulated by the antenna response functions of detectors. By combining the \gls{GW} waveform with the antenna response function of TianQin, we can derive the corresponding \gls{GW} signals that include the response in the frequency domain, as presented in the analytical formulas in \cite{liu20}. 
In Fig. \ref{fig:gw signal}, we plot the sky-polarization-averaged \gls{GW} signals for \glspl{IMBBH} formed in the simulations. This indicates that only inspiral \gls{GW} signals would be observed by space detectors such as TianQin and LISA.

\subsection{Signal-to-noise ratio}
The strength of signals in the data recorded by detectors can be characterized by \gls{S/N}. The optimal S/N of signal $h$ accumulated in observation time in one detector can be calculated by
\begin{align}\label{eq:S/N}
    \rho=\sqrt{(h|h)},
\end{align}
where the inner product between two signals $h_{1}$ and $h_{2}$ is defined as 
\begin{align}
    (h_{1}|h_{2})=4\mathfrak{R}\int_{f_{\rm inital}}^{f_{\rm final}}\frac{\tilde{h}_{1}^{*}(f)\tilde{h}_{2}(f)}{S_{n}(f)}df.
\end{align}
The symbols $h_{1}(f)$ and $h_{2}(f)$ are the signals in frequency domain, and $^{*}$ represents their complex conjugate. The one-sided power spectral density (PSD) of detector noise is denoted by $S_{n}(f)$ in the frequency domain. The initial and final frequencies of inspiral \glspl{GW}, $f_{\rm initial}$ and $f_{\rm final}$, can be derived by
\begin{align}
    f(t)=\left(\frac{5}{256}\right)^{3/8}\pi^{-1}\mathcal{M}^{-5/8}(t_{c}-t)^{-3/8},
\end{align}
where the chirp mass $\mathcal{M}=(m_{1}m_{2})^{3/5}/(m_{1}+m_{2})^{1/5}$. The observation time of detectors and merger time of BBHs are represented by $t$ and $t_{c}$, respectively. Note that for IMBBHs merging within the 5-year observation period, we applied a cutoff for the final frequency as 1Hz. For the case where one signal is observed by multiple detectors, the total S/N is
\begin{align}
    \rho=\sqrt{\sum_{j=1}^{n} \rho_j^2},
\end{align}
where $\rho_{j}$ is the S/N of signal in the $j$th detector. We adopt the noise PSD model of TianQin and LISA from \cite{liu20} and \cite{robson19}, respectively.
It is worth mentioning that the cosmic redshift $z$ and the mass parameters are totally degenerate.
Therefore, in practice, we used the redshifted chirp mass $\mathcal{M}_z= (1+z)\mathcal{M}$.

\subsection{Fisher information matrix}

Assuming the noise of detectors is Gaussian and stationary, the measurement precision of physical parameters can be approximated by the Fisher information matrix (FIM) method
\begin{align}
    \Gamma_{ij}=\bigg(\dfrac{\partial h}{\partial \theta^i}\bigg|\dfrac{\partial h}{\partial \theta^j}\bigg),
\end{align}
where $\theta$ is the physical parameter set which determines the \gls{GW} response signals. For multiple detectors, their total FIM is simply the summation of the FIMs of individual detectors
\begin{align}
    \Gamma_{ij} = \sum_{k=1}^{n}\Gamma_{ij}^k,
\end{align}
where $\Gamma_{ij}^k$ is the FIM of $k$th detectors. As the Cramér-Rao Lower Bound (CRLB) states, the covariance matrix $\Sigma$ of any unbiased estimator is lower-bounded by the inverse of the $\Gamma$, i.e., $\Sigma=\Gamma^{-1}$. The measurement precision of $\theta^{i}$ is the square root of $ii$ component of covariance matrix $\Sigma$.

\section{Result}\label{sec:result}

In this section, we estimate the detection capacity of TianQin for \glspl{IMBBH} in Pop III clusters, including detection horizon, detection number, and parameter measurement precision, under the assumption that TianQin will operate for 5 years with ``3 months on + 3 months off'' observation scenario.

\subsection{Detection horizon distance}

\begin{figure}[hbt!]
\raggedright
\includegraphics[width=1\linewidth]{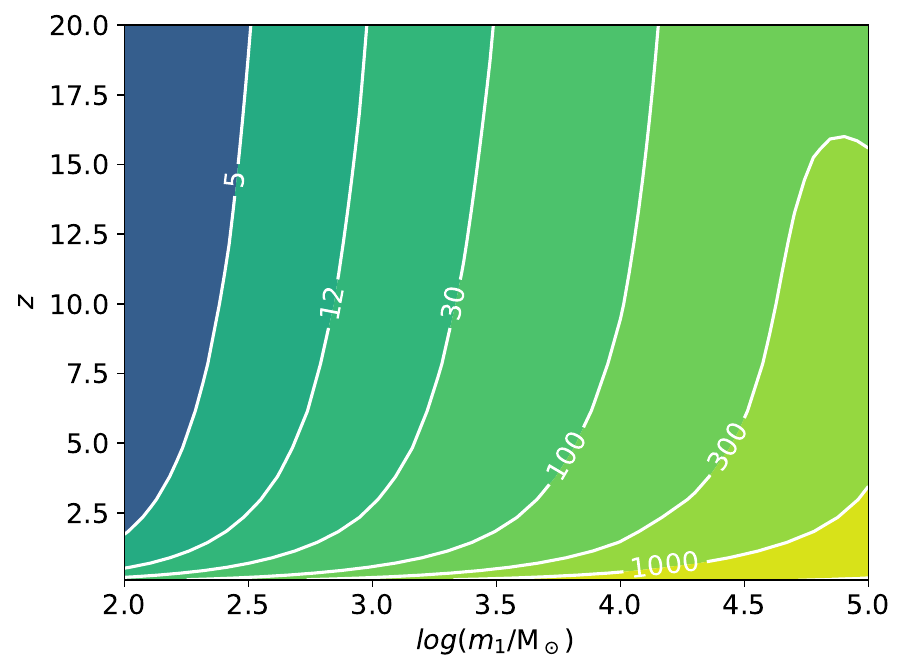}
\caption{Horizon distances for face-on, equal-mass IMBBHs merging within 5 years, located along TianQin’s normal vector ($\bar{\theta}_{S}=94.7^\circ$, $\bar{\phi}_{S}=120.44^\circ$). Masses are given in the source frame.}
\label{fig:horizon}
\end{figure}

The maximum distance that TianQin could observe, i.e., horizon distance, can be obtained by solving Eq. \ref{eq:S/N} with a given S/N, as shown in Fig. \ref{fig:horizon}. 
When S/N=5, TianQin would detect IMBBHs across a broad mass range, extending up to $z=20$. 
As the S/N increases, the mass range of detectable IMBBHs shifts toward higher masses. 
This is because only the stronger GW signals from heavier IMBBHs could accumulate enough to reach the given S/N.
However, when S/N reaches a very high value, e.g., S/N=300, TianQin could not detect IMBBHs at very high redshift, because GW signals from distant IMBBHs are too weak to accumulate enough S/N. 
For IMBBHs formed in Pop III clusters falling in the mass range between $\mathcal{O}(10^{2})M_{\odot}$ and $\mathcal{O}(10^{3})M_{\odot}$, as shown in Fig. \ref{fig:mass distribution}, TianQin could detect sources up to $z=20$ when S/N=5, even as S/N increases to 12, TianQin would still detect sources up to $z=15$. 
This indicates that TianQin has the excellent capability of observing IMBBH mergers throughout the majority of the observable universe.

\subsection{Detection number}

\begin{figure}[hbt!]
\centering
\includegraphics[width=\linewidth]{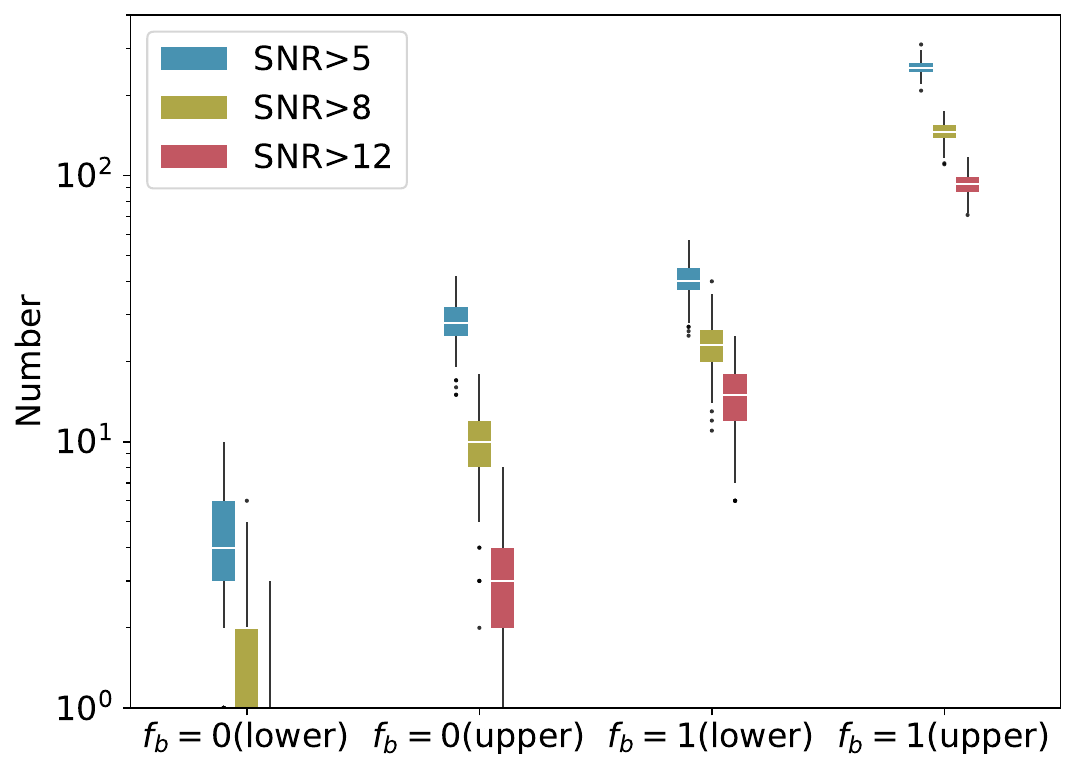}
\caption{Box plots of IMBBH detection numbers from 200 catalogs for $f_{\rm b}=0$ and $f_{\rm b}=1$, under lower and upper merger rates. Medians are shown as horizontal lines within the boxes.}
\label{fig:det-num}
\end{figure}

For each scenario of $f_{\rm b}$, 200 \gls{IMBBH} source catalogs were generated based on the underlying \gls{IMBBH} population. TianQin’s detection performance was then analyzed for these catalogs, with the results presented in Fig.\ref{fig:det-num} and Table \ref{tab:det-num}.

The resulting average number of detections reveals significant variations depending on the S/N threshold, $f_{\rm b}$, and merger rate assumptions over a 5-year operation time.
The upper and lower correspond to the upper and lower limit of merger rate of IMBBHs listed in Table \ref{tab:merger_rate}. 
The average detection numbers range from 1(lower) for the $f_{\rm b}=0$ scenario with applying an S/N threshold of 12, as the most pessimistic result, to 253(upper) under the $f_{\rm b}=1$ scenario for an S/N threshold of 5, as the most optimistic result.
Readers are reminded that the accumulation of S/N is non-linear over time, so the detection number does not scale with observation time linearly.

\begin{table}[hbt!]
\centering
\resizebox{\linewidth}{!}{
\begin{tabular}{lllll}
 \toprule
 Situation & $f_{\rm b}=0$(lower)&$f_{\rm b}=0$(upper)&$f_{\rm b}=1$(lower)&$f_{\rm b}=1$(upper)\\ \toprule
 $\bar N$(S/N>5) & 5& 29& 41& 253\\ \midrule
 $\bar N$(S/N>8) & 2& 10& 23& 146\\ \midrule
 $\bar N$(S/N>12) & 1& 3& 15& 93\\ \bottomrule
\end{tabular}
}
\caption{\label{tab:det-num}The average detection number for 5 year observation.}
\end{table}

Approximately 10\% of IMBBHs exhibit eccentricities above 0.8. To investigate the effect of eccentricity, we computed the dominant harmonic and frequency, following approaches from \cite{wen03} and \cite{hamers21}, for IMBBHs from 168 N-body simulations. Among 39 IMBBHs from the $f_\text{b}=0$ samples and 199 IMBBHs from the $f_\text{b}=1$ samples, only one system has an initial dominant frequency exceeding 1Hz, while nearly 74\% of the $f_\text{b}=0$ systems and 62\% of the $f_\text{b}=1$ systems have an initial dominant frequency above 0.01Hz. These results indicate that although eccentricity increases the dominant frequency, most systems remain within the TianQin band.

\begin{figure}[hbt!]
\centering
\includegraphics[width=1\linewidth]{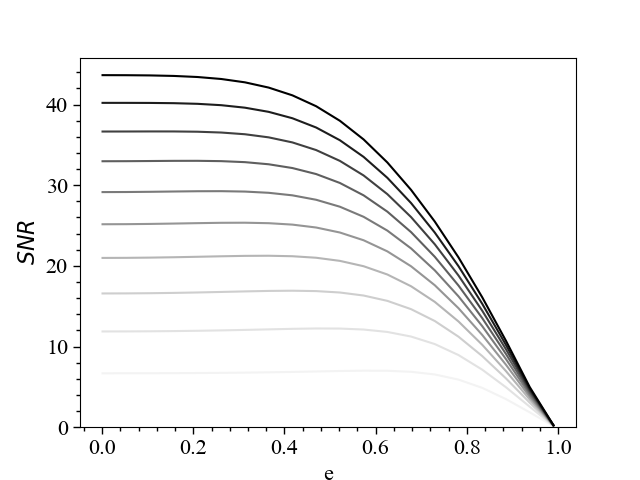}
\caption{S/N variation with eccentricity for IMBBHs of equal-mass components and fixed semi-major axis (0.005 AU). Gray levels indicate masses from $100 M_\odot$ to $1000M_\odot$.}
\label{fig:S/N-ecc}
\end{figure}

Furthermore, we calculated the \gls{S/N} for \glspl{IMBBH} with different eccentricities and component masses from $100M_\odot$ to $1000M_\odot$ based on methods from \cite{peters64} and \cite{ kremer19}. Fig.\ref{fig:S/N-ecc} illustrates how eccentricity reduces the S/N: for $e=0.6 (0.8, 0.9)$ the reduction is approximately 0–20\% (20–60\%, 70–90\%). For instance, a detectable IMBBH with parameters  $m_1=244M_\odot,~ m_2=230M_\odot, a=0.005\text{AU}, e=0.9, z=0.64$ experiences an S/N reduction of about 6\% (40\%, 70\%) when $e=0.6 (0.8, 0.9)$.

Therefore, for sources with high eccentricity ($e>0.6$), our S/N estimates are likely overestimated, and the total number of detectable sources should be lower than those listed in Table \ref{tab:det-num}. Nevertheless, even with an eccentricity of $e=0.8$, systems originally detectable with $\text{S/N}>12$ would still likely exhibit $\text{S/N}>5$. Therefore, future TianQin detections could serve as an important channel to reveal the formation and evolution of IMBHs as well as \gls{Pop III} clusters.

\subsection{Fisher uncertainty}
\begin{figure*}[hbt!]
\centering
\includegraphics[width=0.9\textwidth]{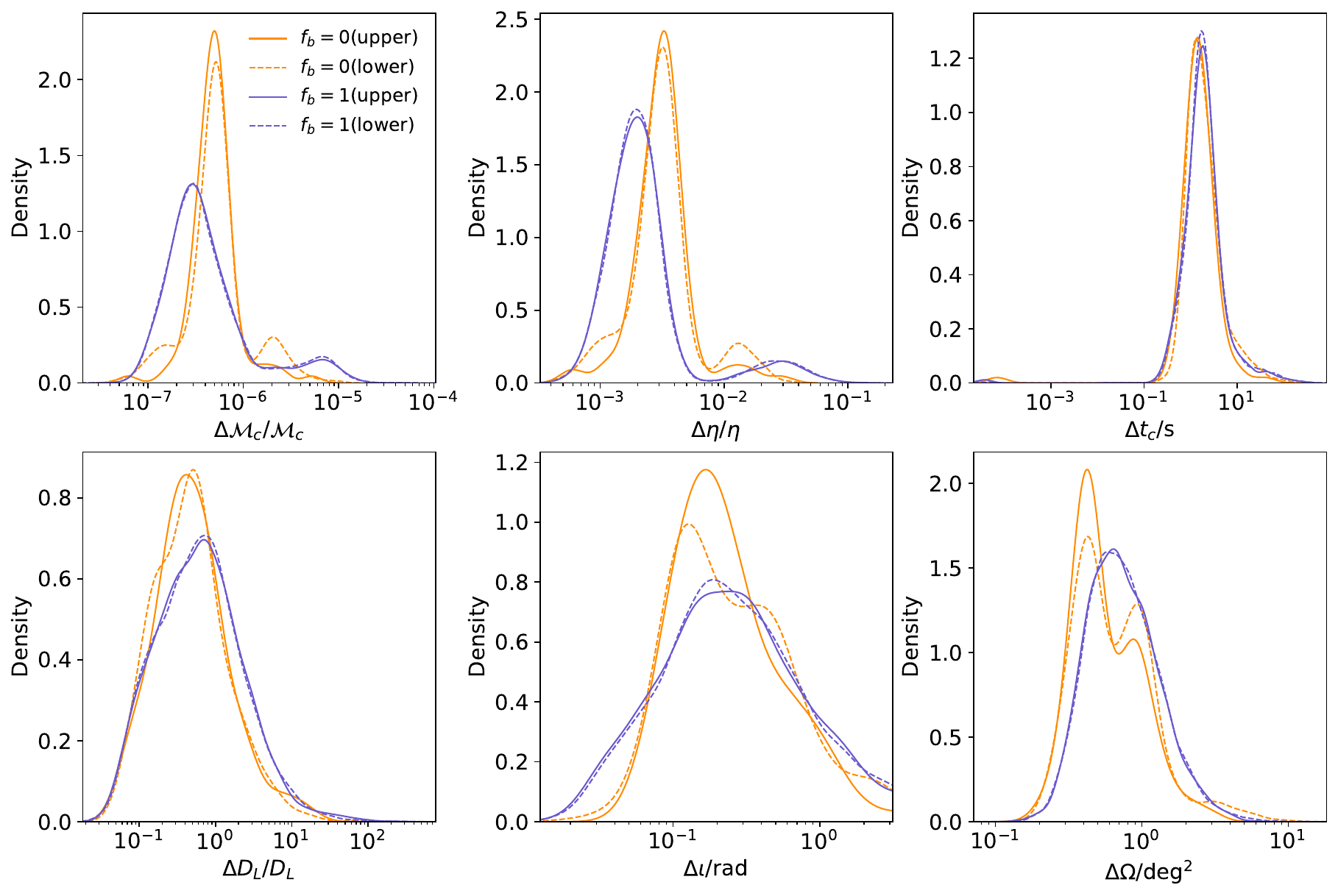}
\caption{Fisher uncertainties for IMBBHs in Pop III clusters with S/N ≥ 12. Blue (orange) lines represent $f_{\rm b}=0$ (1); solid (dashed) lines indicate upper (lower) merger rates. Smoothed distributions are shown for $\mathcal{M}_c$, $\eta$, $D_L$, $t_c$, $\iota$, and $\bar{\Omega}_S$.}
\label{fig:PE precision}
\end{figure*}

For those detectable sources with S/N>12, we further calculated the Fisher uncertainty for four scenarios with varying primordial binary fractions and merger rates. 
We present the results in Fig. \ref{fig:PE precision}, shown with the probability density distributions of relative error for the chirp mass $ \Delta \mathcal{M}_c / \mathcal{M}_c $, symmetric mass ratio $ \Delta \eta/\eta $, and luminosity distance $ \Delta D_L / D_L $, as well as absolute error for coalescence time $ \Delta t_c $,  inclination angle $ \Delta \iota$, and sky localization $ \Delta \bar{\Omega}_S $, which is calculated by
$$
\begin{gathered}
\Delta \bar{\Omega}_S=2 \pi\left|\sin \bar{\theta}_S\right|\left(\Sigma_{\bar{\theta}_S \bar{\theta}_S} \Sigma_{\bar{\phi}_S \bar{\phi}_S}-\Sigma_{\bar{\theta}_S \bar{\phi}_S}^2\right)^{1 / 2}.\\
\end{gathered}
$$

We can first observe that the distribution of different rates (indicated by solid and dashed lines) shows no significant difference. 
This is expected as the rate should only affect the sample size. 
Meanwhile, the primordial binary fraction does change the overall precision distributions, although the difference is not significant and can go both directions.
For the mass parameters, higher $f_{\rm b}$ corresponds to a more precise constraint. 
This can be explained by the fact that the $f_{\rm b}=1$ case has a narrower peak at the lower end of the IMBBH mass spectrum, as can be observed in Fig. \ref{fig:mass distribution}.
Roughly speaking, the precision of mass parameters is inversely proportional to the number of cycles that a source spends in the band.
With a fixed observation time, the lighter sources are associated with higher frequencies, thus a higher number of cycles and better constraints.
This behavior is also consistent with previous studies of on stellar-mass binary black holes \citep{liu20} and massive binary black holes \citep{wang19}.
The precision on merger time is much less sensitive to the $f_{\rm b}$ parameter.
This is also reflected in the relevant studies.
The constraints on the luminosity distance are poor, with the relative error in $D_L$ exceeding 100\% in nearly half of the cases.
This behavior is also observed in previous studies of GW190521-like events \citep{liu22}.
We conclude that the same logic applies to this study: the $D_L-\iota$ degeneracy makes it challenging to pinpoint either parameter.
This can also explain the spread of the $\iota$ distribution extending to the physical limit of $\pi$.
The difference in sky localization is less obvious and shows a different tendency, that $f_{\rm b}=0$ case instead shows a more precise constraint, which is also consistent with previous studies.

Even a more pessimistic result with $ f_b = 0 $ shows, that TianQin can constrain the chirp mass $ \mathcal{M}_c $ with a relative uncertainty of $10^{-6}$, and determine the symmetric mass ratio within $10^{-2} $ relative error. Meanwhile, TianQin can constrain the error of coalescence time $t_c$ within 1s and for the sky localization $\bar{\Omega}_S $ within 1 deg$^2$.
The result indicates TianQin's ability to constrain the mass and location of IMBBHs with high precision.
On the other hand, TianQin could not determine the distance accurately with a high uncertainty for luminosity distance $D_L$ with $\Delta D_L/ D_L$ around 1.
Also for the inclination angle $ \Delta \iota $, the relative error is higher than 0.1. 

Overall, it is shown that for IMBBH events, TianQin can estimate the mass, sky location, and merger time well while it is less capable of estimating the distance and inclination angle.

\section{Conclusion}\label{sec:conclusion}
IMBHs have been a key target for both theoretical and observational astrophysics.
The successful identification and reliable measurement of their parameters can provide pivotal insight into the formation history and their co-evolution of the environment. 
Pop III star clusters, theorized to exist in the metal-poor dark matter halos, could be an important site for producing IMBHs.
Recent studies reveal that IMBHs can form binaries and later merge \citep{wang22}. 
In this work, we extend the direct N-body simulation of Pop III clusters to the whole universe, and by investigating the simulated IMBBH mergers, we present the expected gravitational wave detection ability with the TianQin observatory.
We further calculated the expected Fisher uncertainties. We find that even in the most pessimistic scenario, and with an S/N threshold of 12, TianQin is still capable of observing IMBBH mergers. The intrinsic parameters that determine the phase evolution can usually be very precisely determined, and the relative uncertainty of chirp mass can be determined to the level of $10^{-6}$.
Meanwhile, due to the degeneracy between luminosity distance and inclination angle, neither of these parameters can be precisely determined. The overall behavior of the parameter precisions is consistent with previous studies. 

As indicated by Fig. \ref{fig:gw signal}, the IMR signal of IMBBH can cover both the mHz and the hundred Hz range, making them important potential sources for multi-band GW observations.
Therefore, it is interesting to compare the detection abilities among different GW detectors.
Ground-based detectors can detect \glspl{IMBBH} with masses ranging from $100-10^3 M_\odot$, while spaceborne detectors are more sensitive to the mass ranges of $10^3-10^5 M_\odot$. 
LISA can possibly detect \glspl{IMBBH} up to redshift z = 20 with S/N up to 100 \citep{amaro17}. 
Combinations of next-generation ground-based detectors such as \gls{CE} and \gls{ET} could probe \glspl{IMBBH} with component masses around $1000 M_\odot$, a 3-detector network can achieve mass measurement errors of $\lesssim 0.1\%$ at $z=0.5$ and $\lesssim 1\%$ at $z=2$ with high accuracy of redshift measurement, localized within 1deg$^2$ with $z\lesssim0.5$. For binaries with $m_{1,2}\lesssim 300M_\odot$, the redshift can be measured with $\mathcal{O}(10)\%$ accuracy up to $z=10$, which is precise enough for potential electromagnetic counterpart searches \citep{reali24}. 
The atom interferometers such as AION, ZAIGA, and AEDGE would also be capable of detecting \glspl{IMBBH} \citep{torres23}. DECIGO could constrain the masses and spins of the \glspl{IMBH} within 10\% and issue alerts $10^2-10^3$ s before the coalescence, so that the ground-based detectors can be prepared for the observation of the ringdown \citep{yagi12}.

Our analysis has a number of caveats.
To start with, we extrapolated the evolution of IMBBHs in Pop III clusters from 168 direct simulations.
The extrapolation relies on several assumptions, such as the mass of a typical Pop III cluster, the discrete value of the primordial binary fraction, etc.
So even though the predicted detection number can cover two orders of magnitude, in reality, the scatter could be larger.
Also, the adopted waveform is only reliable for close mass ratio, quasi-circular, aligned-spin binary black holes, while some of the discussed sources can have large mass ratios, non-negligible eccentricities, and randomly oriented spins.
Although we expect that the Fisher uncertainty should be correct in terms of order-of-magnitude, and the S/N estimation should be less severely affected, more accurate estimation would need a more accurate waveform model that is reliable in a larger parameter space. 

In the future, we plan to extend the analysis to a wider range. 
In this study, we only consider the \glspl{IMBBH} in Pop III star clusters, while \glspl{IMBBH} can form in a variety of astrophysical environments. 
Possible alternative locations include globular clusters, where the high stellar density can lead to frequent dynamical interactions, promoting the formation and merger of \glspl{IMBBH}. 
In addition, active galactic nuclei (AGNs) disks can serve as a fertile ground for IMBBH formation and mergers. The presence of massive gas inflows and interactions with the disk environment could drive the pairing and eventual coalescence of \glspl{IMBBH}.
Dwarf galaxies also provide potential hosts for \glspl{IMBBH}, particularly due to their lower stellar mass and the possibility that they retain relic \glspl{IMBH} from earlier cosmic epochs. These smaller galactic systems may hold clues to the early formation pathways of black holes and their role in hierarchical galaxy formation.
The extension to a wider array of formation venues can reveal a more comprehensive view of the formation and evolution of IMBHs.
The future detailed studies could also provide distinctive features for IMBBHs from different channels, better preparing us for using TianQin observations to probe the underlying environment as well as their formation track.

The other possible extension is to further explore the potential of multi-band GW observations.
TianQin can observe the IMBBH inspiral for years in the milli-Hertz band, before finally merge and become detectable for ground-based high-frequency GW detectors.
The coordination of detectors operating in different bands at different times has the potential to provide stringent constraint on gravity theory, or enable us to examine the immediate environment of IMBBHs in details.
To unleash the full potential of this multi-band collaboration, a more detailed studies is warranted. 

\begin{acknowledgements}
This work has been supported by the National Key Research and Development Program of China (No. 2023YFC2206701) and the Natural Science Foundation of China (Grants  No.  12173104, No. 12261131504). Shuai Liu thanks the support from Zhaoqing City Science and Technology Innovation Guidance Project (No. 241216104168995) and the Young Faculty Research Funding Project of Zhaoqing University (No. qn202518). L.W. thanks the support from the National Natural Science Foundation of China through grant 21BAA00619 and 12233013, the High-level Youth Talent Project (Provincial Financial Allocation) through the grant 2023HYSPT0706, the one-hundred-talent project of Sun Yat-sen University, the Fundamental Research Funds for the Central Universities, Sun Yat-sen University (22hytd09).
We also thank Jian-dong Zhang, En-kun Li, Xiang-Yu Lyu, Lu Wang and and Jian-Wei Mei for their useful comments. The authors acknowledge the uses of calculating utilities of 
\texttt{Numpy} \citep{van11}, \texttt{Pandas} \citep{mckinney11} , and \texttt{Matplotlib} \citep{hunter07} for data analysis and plotting. 
\end{acknowledgements}

\bibliographystyle{aa} 
\bibliography{reference}  

\end{CJK*}
\end{document}